# A Survey of Multimedia Streaming in LTE Cellular Networks

Ahmed Ahmedin*, Amitabha Ghosh†, and Dipak Ghosal*
*University of California, Davis, CA 95616
{ahmedin, dghosal}@ucdavis.edu
†UtopiaCompression Corporation, Los Angeles, CA 90064
amitabhg@utopiacompression.com*Abstract*—With the growing of Long Term Evolution (LTE) cellular networks and the increase in the demand of the video services, it is vital to consider the challenges in the streaming services from a different perspective. A perspective that focuses on the streaming services in light of cellular networks challenges, both per layer basis and across multiple layers as well. In this tutorial, we highlight the main challenges that faces the industry of video streaming in the context of cellular networks with a focus on LTE. We also discuss proposed solutions for these challenges while highlighting the limitations of these solutions and the conditions/assumptions required for these solution to deliver high performance. In addition, we show different work in cross layer optimization for video streaming and how it leads towards a more optimized end to end LTE networking for video streaming. Finally, we suggest different open research areas in the domain of video delivery over LTE networks that can significantly enhance the quality of streaming experience to the end user.## I. INTRODUCTION

Video streaming is currently one of the fastest and most expending services due to the emergence of multimedia based applications. The nature of these applications varies between business video conferences and telesurgeries to home entertainment, including but not limited to security surveillance and tracking operations. The evolution in wireless networks adds the mobility as another dimension to the streaming services. The high bit rates demands by users impose challenges on the network operators and vendors to continuously develop and enhance the cellular networks capabilities which leads to creating new services in addition to enhancing the quality of the existing ones. Furthermore, today's mobile devices (e.g., iPhone, iPad, Android, tablet) not only have advanced capabilities in terms of more processing power, longer battery life, higher resolution display, and a variety of form factors, but also support seamless execution of numerous applications (Apps) developed by third parties. This expansion of smart mobile devices industry is driving service providers and operators to introduce more effective techniques to bring high-quality services for the end users. Video streaming services are one of the services that was highly affected by the cellular networks and industry evolution. Content providers such as YouTube and Netflix direct some of their potential towards perfecting mobile apps and enhancing the streaming quality on mobile devices. As shown in Figure 1, Reelseo, a media market guide, claims that video streaming applications represents 35% of the cellular data traffic. It is expected that 70% of the cellular data traffic will be from video by 2016 [1].

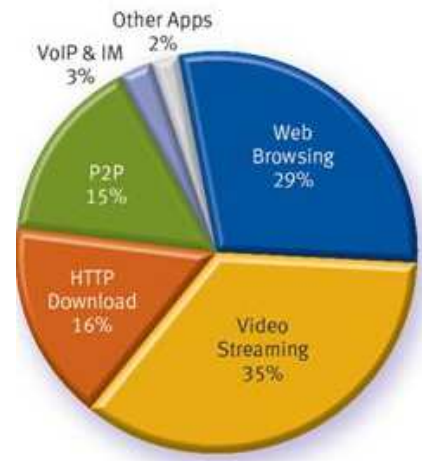

Fig. 1: A study by Reelseo shows that 35% of the cellular data traffic is for video streaming.

However, there are key challenges that affects the quality of video streaming over a cellular network, such as mobility, changing wireless environments, diversity in devices capabilities, power management, the strict delay requirements for the video traffic and other difficulties that have to be considered for successful video transmission. Indeed, increasing the bandwidth and the bitrate can solve some of these problems, however smart and advanced protocols are needed to manage this bandwidth and distribute it fairly among the multiple users and handle different requests. Long Term Evolution (LTE), often referred to as (4G), is one of the promising cellular technologies that has highly evolved, and been developed and deployed over the past few years. The fact that LTE advanced has a peak download data rate of 1 Gbps and upload rate of 0.5 Gbps [2] promotes it as a good candidate for multimedia streaming. Hence, the focus of this survey is to shed some light over the mechanisms of video delivery over wireless cellular networks in general and LTE in particular.

We approach the challenges of the video streaming problem on a per layer basis and we specifically focus on LTE networks. We discuss the main challenges for each layer and some proposed solutions for each problem in addition to pointing out the limitations of these solutions and the conditions needed for them to perform well. The survey also propose some open research areas for each layer. In addition, we survey some cross layer optimization solutions, where interaction/coupling happens between two or more layers. Unfortunately, quantifying the effect of the solutions on the video quality is very difficult due to the existence many performance metrics such as frame delay, peak signal to noise ratio, among others that are hard to connect together [3] and each focus on a different vital aspect of the video performance. Hence, we show the impact of each solution on the different streaming performance aspects.

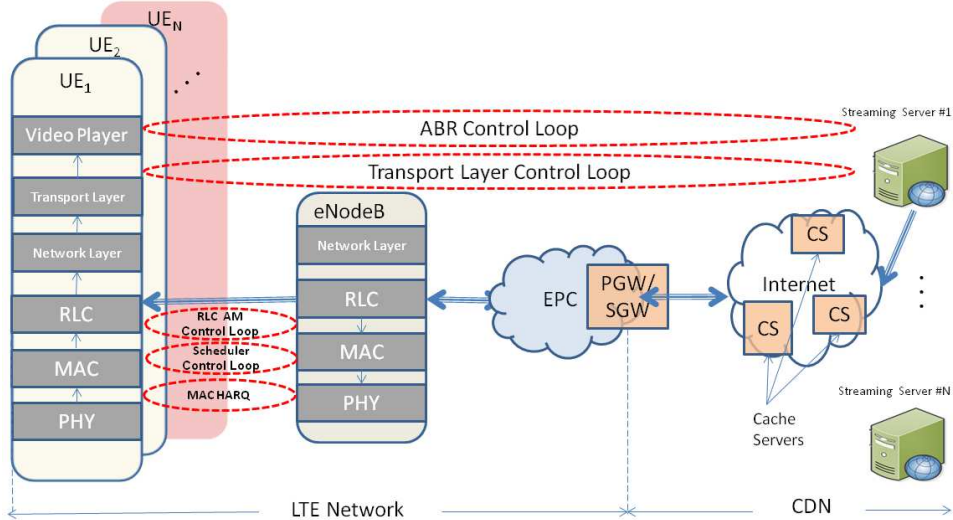

Fig. 2: End to end full system architecture including LTE radio network, LTE EPC network and the content distribution network

There has been many related surveys addressing the multimedia streaming problems over wireless networks in general such as Zhu and Girod [4] which provides an overview of the technical challenges of video streaming over different types of wireless networks. Mantzouratos *et al.* [5] points out the suitability of cross layer design for optimizing video streaming over mobile ad hoc networks (MANETs). A survey of issues in supporting QoS in MANETs was presented by Mohapatra in [6]. However, none compare between the pros and cons of the different existing approaches and also there is no focus on the cellular technology specially 3G and 4G. Our tutorial is considered as a supplementary to the existing surveys as we address the video streaming solutions proposed to overcome the problems of each layer starting from the application layer to the medium access control (MAC). We also focus more on the LTE technology and identify the advantages, the disadvantages and the limitations for each of the discussed solutions. We discuss the advantage of cross layer design and show some successful approaches in the cross layer design context while providing directions for open problems and future research.

The rest of this survey is organized as follows: In section II, we introduce an overview about the LTE networks. Section III gives an overview over the structure of the video coding types used for streaming and discusses some famous parameters to quantify the video quality. Sections IV, V, and VI discusses respectively the application, transport and MAC layers problems and possible solutions. A cross layer approaches are discussed in Section VII. Finally, we conclude our tutorial in Section VIII.

## II. LTE Overview

The LTE project started in 2004 by the Third Generation Partnership Project (3GPP) telecommunication body to enhance the cellular communication. It started as an evolution from the Universal Mobile Telecommunication System (UMTS). LTE Advanced is expected to achieve peak downlink rates up to 1Gbps [2]. Besides the high data rate and the wide coverage range that LTE provides, it is backward compatible with the previous cellular networks generations. There has been some competitors to the LTE systems such as WiMAX defined by the standard IEEE 802.16e which provides high data rates and mobility advantage similar to the LTE. However, the fact that LTE supports seamless connection to existing cellular networks and the simple compatible architecture, which reduces the operating expenditure (OPEX), promotes LTE to be the perfect candidate for the next generation of cellular networks [7].

### A. LTE Architecture

LTE has been designed to support the packet switching services. Hence, the system architecture has been devolved to contain the Evolved Packet System (EPS) in addition to the regular connectivity radio core network functions. In this section, we will show some of the related EPS functionality to the video streaming and give an overview for the core network.

Figure 2 shows the network architecture, including the LTE radio network, LTE core network and the content distribution network (CDN). These three parts provide the end to end connectivity between the user equipments (UEs) and the content provider. The eNodeB is considered the radio access part of the LTE network, which provides the radio connectivity to the UE. The figure also highlights the different control loops which ensures smooth end to end video delivery and will be discussed later in depth throughout the entire survey.

The physical layer of an LTE downlink uses orthogonal frequency-division multiple access (OFDMA), and allocates radio resources in both time and frequency domains, as shown in Figure 3. The time domain is divided into LTE downlink frames, which are split into Transmission Time Intervals (TTIs) of duration 1 millisecond (ms) each. The LTE downlink frame has a total duration of 10 ms corresponding to ten TTIs. Each TTI is further subdivided into two time slots, each of duration 0.5 ms. Each of these slots corresponds to 7 OFDM symbols. In the frequency domain, the available bandwidth is divided into subchannels of bandwidth 180 kHz each, and each subchannel comprises 12 adjacent OFDM subcarriers. As the basic time-frequency unit in the scheduler, a single Physical Resource Block (PRB) consists of one 0.5 ms time slot and one subchannel. The minimum unit of assignment for a UE is one PRB, and each one can be assigned to only a single UE. Additionally, the LTE downlink makes use of adaptive modulation and coding (AMC) to match the transmission parameters to changing wireless channel conditions. In AMC, the modulation and the coding changes based on the wireless environment to provide more robustness in weak channels and higher data rates over the strong PRBs.

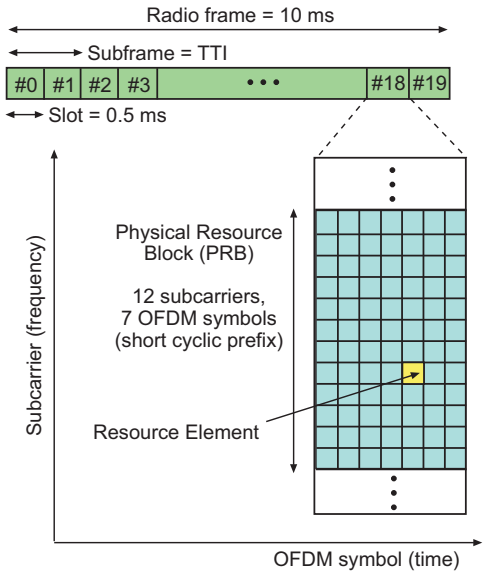

Fig. 3: LTE downlink frame structure showing a physical resource block (PRB), a resource element, and the durations for a time slot, a subframe, and a frame. A PRB consists of 12 consecutive subcarriers and 7 or 6 OFDM symbols depending on whether a long or short cyclic prefix is used, respectively. The basic time-frequency unit in the scheduler is a single PRB.

## III. VIDEO STREAMING OVERVIEW

A video is considered as a time sequence of still-images. As the wireless communication technology are becoming more advanced, the video streaming services are becoming more sophisticated and cover wide range of applications. These applications vary from entertainment purposes such as video on demand, video chatting and interactive gaming to business purposes such as tele-surveillance, video conferencing and remote learning. Every one of these applications has its own requirements and performance metrics to emphasize. For example, the effect of few seconds lagging is not severely sensitive for video on demand services, however it is very critical for video conferencing.

Al-Mualla and Canagarajah in [8] introduce the main challenges for video streaming over cellular networks. One of the main challenges in the cellular networks is the limited bandwidth, hence efficient video compression techniques are needed to overcome the bandwidth problem. Also, compression techniques with higher compression ratios are needed to enhance the video quality by fitting more frames and details into the same container then the radio spectrum. Another challenge is the different capabilities of the UEs. Mobile devices range from battery and hardware constrained cell phones, to more powerful tablets with sophisticated transcoding features. Hence, video codecs implementations over cell phone need to take into consideration the computation complexity and the limited battery life for cellular devices. Until 2000, implementations of video codecs [9], [10] indicate that digital signal processors (DSPs) could not achieve real-time video encoding. Recently, there is more focus on low complexity codecs implementation on embedded processors such as [11].

In addition to coding challenges, we consider the severity of the mobile channel is the most difficult. The link quality depends the UE's distance from the eNodeB as well as the shadowing and the fading. These effects reduce the reliability of video delivery over the wireless spectrum. Hence, a possible solution is using forward error correction codes shall reduce the effect of the wireless channel degradation, however it introduces extra redundancy to the video packets [12]. These redundancy comes on the expense of the first challenge that we explained due to the limited radio spectrum. To figure out the correct coding, feedback signaling should be used by making the UE report the quality. Unfortunately, it is hard to have a unified metric to quantify the video quality. Hence, a quantitative comparison between different schemes is difficult as each scheme focuses on fixing a certain problem and displays the advantage by showing the effect on the related performance metrics. Furthermore, the objective video quality metrics are built based on mathematical models to quantify the subjective quality assessments that usually needs a trained eye to judge it. For the rest of this section, we will discuss some popular video performance metrics.

### A. Peak Signal to Noise Ratio (PSNR)

PSNR is a full-reference video quality metric, i.e., uses the distortion-free version of the video as the reference. Assuming we have a video $I$, hence PSNR is measured in reference to a video $R$, typically a high quality or a distortion free version of the video $I$. If the frame size is $u \times v$ (in pixels), the PSNR of the $i^{th}$ frame, PSNR(i), can be calculated using the mean square error (MSE) between the $i^{th}$ frame in the video $I$, $I_i$, and its correspondence in the video $R$, $R_i$, as follows [13],

$$\text{MSE}(i) = \frac{1}{uv} \sum_{k=0}^{u-1} \sum_{l=0}^{v-1} [I_i(k,l) - R_i(k,l)]^2, \quad (1)$$

$$\text{PSNR}(i) = 10 \log_{10} \left( \frac{\text{MAX}^2}{\text{MSE}(i)} \right), \quad (2)$$

where MAX is the maximum possible pixel value (typically, 255). Hence, the average video PSNR is given as,

$$\text{PSNR} = \frac{\sum_{i=0}^{z} \text{PSNR}(i)}{z}, \quad (3)$$

where $z$ is the number of frames in $I$ and $R$. PSNR is the most widely used objective video quality metric because of its simplicity. However, PSNR values do not perfectly correlate with the perceived visual quality due to the non-linear behavior of the human visual system [14].

### B. Structure Similarity Index Matrix (SSIM)

SSIM is another full referenced metric to characterize the video quality. Since it takes into consideration the interdependency between pixels, it is more consistent with the human eye [15]. SSIM($i$) is calculated on various windows $x$ and $y$ of the frames $I_i$ and $R_i$, respectively.

$$\text{SSIM}_{x,y}(i) = \frac{(2\mu_x \mu_y + c_1)(2\sigma_{xy} + c_2)}{(\mu_x^2 + \mu_y^2 + c_1)(\sigma_x^2 + \sigma_y^2 + c_2)}, \quad (4)$$

where $\mu_x$ and $\sigma_x$ are the the mean and variance for window $x$, respectively; likewise, $\mu_y$ and $\sigma_y$ are the mean and variance for window $y$. The covariance of $x$ and $y$ is $\sigma_{xy}$. The two variables $c_1$ and $c_2$ are to stabilize the division. Hence, the SSIM is given as

$$\text{SSIM} = \frac{\sum_{i=0}^{z} \text{SSIM}(i)}{z}. \quad (5)$$

### C. Video Quality Metric (VQM)

VQM is another objective full referenced quality metric developed by the Institute for Telecommunication Science (ITS). It shows a high correlation with the subjective quality tests. The VQM calibrates the video and corrects the temporal and spatial shifts then extract the different quality features. It combines these quality measurements together using a set of linear combination of 7 parameters based on one of the various models defined for the VQM tool such as television, conference, general and PSNR [16].

### D. Kullback-Leibler Divergence (KLD)

This is a reduced reference metric that is designed to predict the quality of distorted images without full information about the video. It is helpful in the real time streaming schemes such as video conference, where a quick feedback is required from the receiver to the transmitter without much information about the original video. The overall distortion, $D$, between the distorted and reference images can be calculated, according to [17], as

$$D = \log_2 \left(1 + \frac{\sum_{k=1}^{K} \left| \hat{d}^k(p^k || q^k) \right|}{D_0} \right) \quad (6)$$

where $K$ is the number of sub-bands, $p^k$ and $q^k$ are the probability density functions of the k-th sub-bands in the reference and distorted images, respectively, and $\hat{d}^k$ is the KLD between $p^k$ and $q^k$, and $D_0$ is a constant used to control the scale of the distortion measure.

### E. Blind Quality Assessment

These are a no reference category matrix, where no information needed about the original video. The blind techniques are very helpful over the cellular network where a bandwidth utilization is required and the network protocols are trying to avoid extra overhead signaling. Chen and Song [18] analyzed the common mobile video impairments and used it as a metric for the video quality. The first is blockiness where the image contains small blocks of a single color. The second is blurriness where the edge of the image is not as sharp as the neighboring pixels. And finally, the noise in the image due to the random variation in the color of the images. The experimental results for this technique shows that this blind estimation techniques gives a close result to the SSIM metric.

## IV. APPLICATION LAYER MECHANISMS FOR VIDEO STREAMING

The application layer is responsible of deciding the appropriate encoding techniques for the video frames based on the application type and requirements. For example, the video on demand services require transmitting high quality videos, however it can tolerate a reasonable amount of delay hence less transmission errors. On the other hand, real time streaming services are more strict and require low delay and jitter. Video compression/ encoding is very important to utilize the bandwidth. It is considered, as we discussed earlier in Section III, a rare resource and the UEs compete over it. According to [8], a row video data of an HDTV will need at least 1.09 Gbits/s to be appropriately received, while a typical HDTV encoded video application over 6 MHz channel needs only 20 Mbits/s. The idea of compression is to get rid of the redundancy in the frames. Redundancies can be a spatial, such as when part of the picture have the same color like a painted wall. The other type of redundancy is the temporal redundancy, such as consecutive frames having the same background. Shannon's lossless coding theorem [19] states that it is impossible to have a lossless compression coding rate less than the source entropy.

The year 1984 marks the birth of the first video coding international, H120 [20], by the ITU-T formerly known as International Telegraph and Telephone Consultative Committee (CCITT, French of: Comité Consultatif International Téléphonique et Télé graphique). The performance of the H120 was remarkable on the spatial resolution, however it had a very poor temporal resolution. After that, different standards has evolved specially the H26x family and the MPEG family. In the rest of this section, we will discuss some of the most used coding schemes in cellular networks and the recent work related to these schemes.

## A. MPEG-4

The first version of the MPEG-4 standard has been finalized in 1998 [21]. MPEG-4 is designed to work across variety of bit rates starting from a few kilobits per second to tens of megabits per second which makes it very suitable for the unstable wireless environment. The concept of profiles was introduced in the MPEG-4, hence it is suitable for different video sources, communication techniques and applications. The most important development in the MPEG-4 is adapting an object based representation techniques, where the scene is coded based on individual objects rather than pixels. As shown in Figure 4, the frame is divided into different video object planes (VOPs) that can be decoded independently and manipulated.

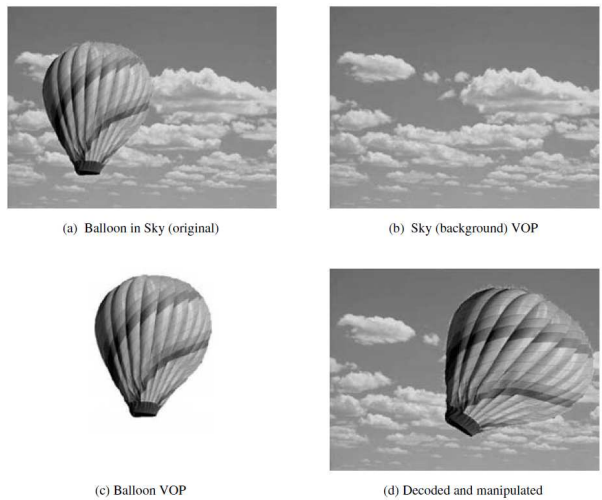

Fig. 4: An example of the object based representation in MPEG-4.

All these advantages encouraged the service providers and the cellular network operators to use MPEG-4 in the different video streaming applications as it satisfies different ranges of requirements. A performance evaluation for the MPEG-4 over the UMTS is shown in [22]. The results show that MPEG-4 over the UMTS in the unacknowledged mode provides timely delivery, but no error recovery. On the other hand, the acknowledged mode enhances the radio link control (RLC) block error rate by $40\%$ with an acceptable video quality due to using the Hybrid Automatic Repeat Request (HARQ) as we will see later in Section VI. Increasing the RLC pre-decoder buffer can help in increasing the video quality and ensuring that the packets are received in a timely fashion. Our research group in [23] explores the use of the MPEG-4 characterstics into novel scheduling techniques to maximize the average quality of a multiple users Wideband Code Division Multiple Access (WCDMA) cellular network. In fact, the network parameters and settings directly affect the pefromance of the MPEG-4 video transmission. These network parameters can be adjusted based on the application as shown in [24], [25]. The MPEG-4 perfromance over LTE is investigated in [26] as it discusses the LTE downlink air interface capacity using realistic MPEG-4 traffic models. The results show the tradeoff between the user outage and the video frame loss for different number of users as shown in Figure 5.

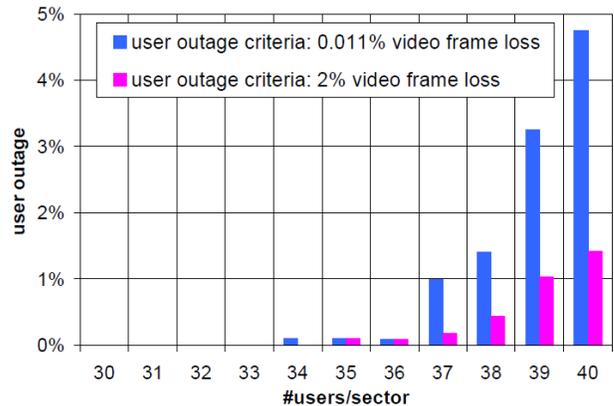

Fig. 5: Tradeoff between user's outage and the frame loss in MPEG-4 over LTE [26].

## B. H.264

H.264, also known as Advanced Video Coding (AVC), is another standard by the ITU that was started in 2003 and completed in 2004. The standard adds many extensions over the MPEG-4 and the H.26x series in general [27]. These extensions can accommodate the new applications requirements, increase the compression ratio and enhance the playback quality. The standard adds five new profiles for professional streaming services, specially real time videos and surveillance applications. A major addition in the H.264 standard is the Scalable Video Coding (SVC). The H.264 SVC standard is well suited for wireless environments in general and the cellular network in particular which exhibit variable link quality due to shadowing, multipath fading, and limited bandwidth [28]. These factors can cause link quality degradation, leading to reduction in the video delivery rate as well as increase in the pixel error rate. SVC grants three different scalability options. The first is a quality scalability, in which, the data and decoded samples of lower quality layers are used to predict higher quality layers to reduce the required UE rate to encode a higher quality layers. The second option is a temporal scalability, where complete frames are dropped from a video by motion dependency. Finally, we have a spatial scalability where videos are coded at multiple resolutions. A streaming device can use any of these scalability options or combine them depending on the type of video, application and user's requirements. Consequently, H.264 SVC has many levels and profiles that differ in the level of compression, bit rate, and size. Another addition in the H.264 is the multi view coding (MVC). This comes in handy to efficiently code the same scene from different viewpoints, hence a frame from a certain camera can be temporally predicted from other cameras' frames in addition to the same camera [29].

All these additions and more made H.264 as one of the most popular video coding schemes for LTE. A statistical and simulation analysis is conducted in [30] to evaluate the H.264 performance over LTE. The authors use some of the metrics discussed in Section III, such as PSNR, SSIM, blocking and blurting, to compare between the different scalability options. However, the results show that scalability by itself is not enough to avoid video quality degradation. Hence, smart scheduling, and efficient routing techniques are needed, which will be discussed in the following sections. The scalability of the SVC, the high LTE data rates and the development of the cloud computing fields add new dimensions to the streaming services specially with social networks. As proposed in [31], cloud based agents are created for each active UE. These agents are responsible of adjusting the video quality using SVC based on the feedback information received from the UE to prefetch the videos from the social networks through the cellular networks. This can help reducing the network congestion by fetching the videos to the users in advance when the network is not crowded such as midnight to 6 am.

### C. High Efficiency Video Coding (HEVC)

The HEVC is the successor of the H.264 in video coding. The standard has been through a lot of development since 2004 until it was finalized in 2013. It is also known by the name of some of its development stages titles such as H.265 and MPEG-DASH [32]. The HVEC aims to increase the compression ratio by two while decreasing the complexity level to half. HEVC has some new specs to increase the compression ratio and the video quality such as coding tree unit (CTU). The HVEC replaces the concept of macroblocks by CTU which is a large block of pixels with variable sizes to better encode the frames. Another extension is the parallel processing ability, where the frame is divided into tiles, and each can be encoded and decoded independently. Also, HEVC has at least four times prediction modes more than the H.264 which makes the prediction more sophisticated and results in better playback quality. Moreover, the HEVC added new profiles to support displays up to $(8192 \times 4320)$ pixels compared to the $(4096 \times 2304)$ pixels in the H.264. An evaluation for the use of HEVC over mobile networks is presented in [33]. In addition, HEVC is considered to be a highly effective encoding standard for some of the recent applications over LTE such as Telemedicine as suggested in [34].

### D. Summary and Discussion

In this section, we discussed different famous video coding schemes that can be suitable for video streaming over LTE networks. It is clear that the video coding industry is always in development and always being pushed to be evolved by different entities. Investments from industry has a huge impact on the direction of the enhancements. Content providers such as Google, Netflix and Hulu are pushing for lower coding complexities without quality degradations. On the other hand, devices manufacturers, such as Sony and Samsung, want the video encoding to introduce a better quality that fully utilizes their hardware platforms to satisfy the end user. As we mentioned earlier, quantitative comparison between different schemes is hard, moreover the tradeoff between the provider and the manufacturer is difficult to characterize. However, there has been some trials to show the compression ratio difference on the same platform between different video schemes shown in [35]. The comparison results are shown in Table I.

| Standard | Average bit rate reduction compared to | | |
|---|---|---|---|
| | H.264 | MPEG-4 | H.263 |
| HVEC | 35.4% | 63.7% | 65.1% |
| H.264 | N.A. | 44.5% | 46.6% |
| MPEF-4 | N.A. | N.A. | 3.9% |

TABLE I: Comparison between different coding schemes based on bitrate reduction [35].

Apparently, HEVC can be the future of the video coding schemes over wireless networks, however HEVC Adoption is still in progress. While HEVC can help content producers, and distributors having a better quality content at the current bitrate, It still needs a lot of work and can form interesting research topics. Also, current hardware platforms upgrades are required, hence compatibility issues which can generate many interesting research problems. Industry investments play an important rule in adopting the HEVC. The current revolutionary movement of using adaptive streaming makes the HEVC a good candidate to be used in video delivery over wireless networks. In general, adaptive video streaming introduces better end to end quality. The content provider adapts their transmitted video quality according to some network measurements, such as congestion, rate of ACKs, and buffer underflow. This adds additional complexity and signaling overhead to the network. This loop of signaling feedback and quality decision represents the Adaptive bitrate (ABR) control loop as shown in Figure 2. Dynamic Adaptive Streaming over HTTP (DASH) is one of the solid examples That uses the ABR control loop concept to enhance the user's experience. DASH is widely used by Netflix, YouTube and Hulu. Using the HTTP protocol, the client downloads chunks from the server then it can seamlessly reconstruct the original media stream. During download, the client dynamically requests fragments with the right encoding bitrate that maximizes the quality of the streaming application , typically determined by factors such as startup delay, video freezes due to re-buffering, and the playback video bitrate [36], [37], and reduce network congestions. There are a number of deployed solutions of DASH. Adobe Dynamic Streaming for Flash [38] is available in latest versions of Flash Player and Flash Media Server which support adaptive bit-rate streaming over the traditional Real-Time Messaging Protocol (RTMP), as well as HTTP. Apple HTTP Adaptive Streaming HTTP Live Streaming (HLS) is an HTTP-based media streaming communications protocol implemented by Apple. Microsoft Smooth Streaming [39] enables adaptive streaming of media to clients over HTTP.

An advanced prototype for streaming using the DASH over LTE is deployed in [40]. The demo results suggest the possibility of adaptively streaming the next generation video standard content over LTE networks with a very reliable quality of experience (QoE). In [41], Thomson Video Networks (TVN) shows the possibility of having a high quality live broadcasting services over LTE, such as HD-TV multi broadcasting for games and events, by combining three technology enablers.

The first is the HEVC for its high compression ratio to save the bandwidth. The second is the dynamic adaptive streaming over HTTP (DASH) to adapt the transmission with the user's channel quality and finally, the evolved Multimedia/Broadcast Multicast Service (eMBMS) to have efficient broadcast delivery to the UEs over LTE.

The problem with such end to end adaptive control loops that it takes long time to respond to the network variations, which affects, sometimes negatively, the rest of the network elements into consideration. In LTE networks, there is fast variations of channel quality and demands. A content provider may decide to assign a certain user a high quality video, according to the measurements, which forces the radio network to either assign more resources to this UE on the expense of other users and other traffic types or the UE may suffer buffer underflow if the network decides not to honor the content provider required rate. Moreover, reporting the LTE network fast pace measurements to the content provider introduces complexity and signaling overhead. Hence, this calls for optimizing all network's elements simultaneously. An alternative to the application layer ABR is to consider a cross-layer design in which the ABR control is tightly integrated with the MAC scheduler. This can open doors to more enhancements and innovative design to the existing schemes. We will discuss this in more details later in Section VII.

## V. TRANSPORT LAYER MECHANISMS FOR VIDEO STREAMING

The transport layer, often referred to as layer 4, is responsible of providing an end to end service for the different applications via the transport layer control loop shown earlier in Figure 2. Transport layer also can provide a reliability option for the received packets by using means of error detection such as checksum then notify the sender using ACK/NAK to retransmit the corrupted or lost packets. In addition, the transport layer applies flow control mechanisms to prevent buffer overflows when the sender is faster than the receiver as well as congestion control mechanisms to mitigate the effect of low quality links and network congestions by slowing down the sender [42]. In multimedia and streaming applications, the transport layer has to ensure the end to end quality and handles many challenges such as jitter, data priorities, packet reordering, delay, bandwidth availability, and session establishing and maintaining [43]. Unlike the wired network, assumptions of no interference can not be applied in wireless networks as packet losses result from the noisy time varying channel as well as the usual congestion reasons [44], [45]. In the following subsections, we show some of the common layer four protocols for video streaming.

### A. Transmission Control Protocol (TCP)

Different transport protocols are used for media streaming. The most well known core transport protocol is the TCP as it supports variety of traffic types including multimedia streaming. The TCP is first specified in 1974 [46]. However, a lot of enhancements and additions are applied to it over the years while keeping the basic operation same. The TCP was originally optimized for wired networks as a connection oriented protocol to handle the flow and congestion control, and ensure the reliability of the received packetized data. Most of the TCP versions have congestion control and reliability assurance mechanisms to retrieve the lost data [47]. The reliability part is obtained using the accumulative acknowledgment scheme where the receiver sends an ACK with a sequence number to inform the receiver of successfully acquiring all the packets perceiving this acknowledged sequence number. The sender retransmits the lost packets. Moreover, TCP uses the ACKs time stamps to get estimates for the round trip time (RTT). Based on the RTT estimated values, the rate of the sender is adjusted to avoid congestions and decrease packet losses.

Unlike wired networks, most of the packet drops in the wireless environments are not from congestions but due to the temporary degradations in the link quality. These losses are due to fading, interference, or shadowing. When packets are lost due to link quality degradation, TCP enters the congestion avoidance mode which avoidably decreases the sender's rate quickly. Consequently, radio resources are wasted due to the sender's rate back off. There has been trials to adapt TCP to the wireless environment and to utilize resources [44], [48]. These modifications contributed in using the TCP in LTE to deliver different data traffic types in general and for media streaming in particular. A study is conducted in [49] to evaluate the performance of the TCP running in an LTE network for a severe vehicular environment. The study shows that the obtained aggregated TCP throughput varies among users based on their radio scheduling algorithms (will be discussed in more details in Section VI) and the channel conditions severity. Although some studies suggest that TCP may not be suitable for media streaming due to the back-off, retransmission mechanisms and delays [50], it is still commonly used in the commercial streaming traffic because of its reliability [51]. An analytical study for the performance of TCP for live and stored media streaming is conducted in [52] under various conditions. The study shows that TCP provides a good performance when the achievable TCP throughout is roughly twice the media bit rate with few seconds of start up delay.

As a result, media streaming over LTE using TCP is possible with slight modifications to optimize the performance of the TCP over LTE networks for video streaming applications. In [53], authors present a novel scheme of adaptive TCP rate control to stream SVC encoded videos to accommodate the varying channel conditions. The TCP rate adaptation scheme adds significant improvement to losses, playback interruption, delay and buffer size. Figure 6, shows the improvements in 3 different measurements when using the TCP rate adaptation along with the SVC. It is worth to mention that the algorithm adopted in [53] is not optimized for LTE, hence a better performance can be obtained by using more optimized TCP versions for LTE.

Another study is conducted in [54] to determine the optimal UE buffer for smooth playback and mitigate the TCP sawtooth throughput behavior. Increasing the receiver buffer leads to smooth playback. However, increasing the receiver buffer for

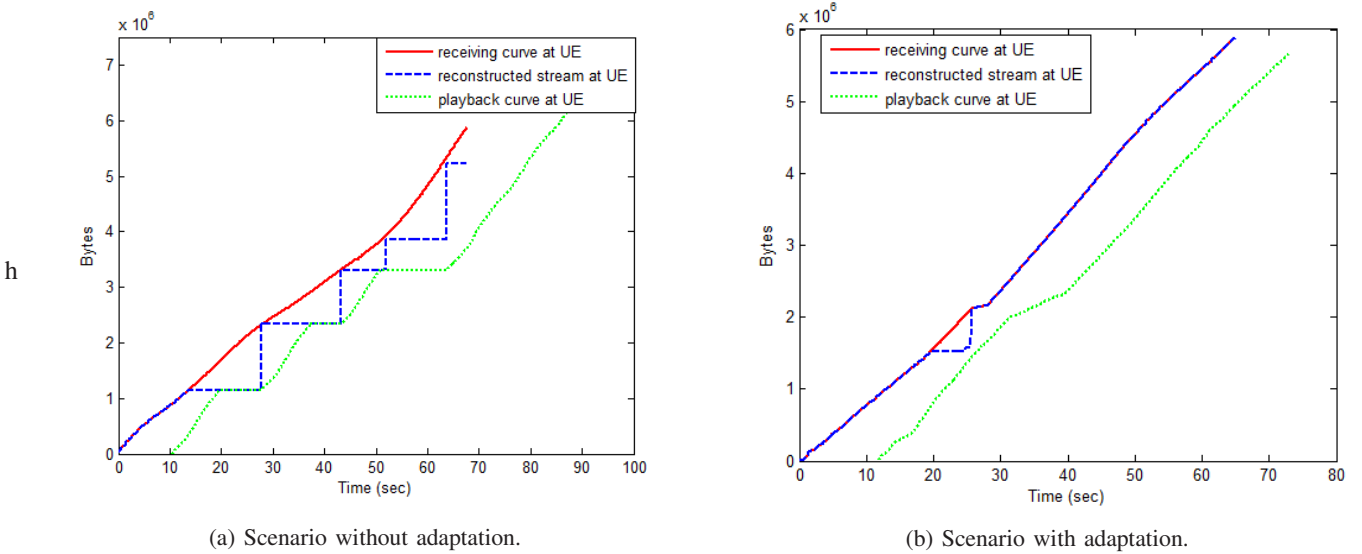

Fig. 6: TCP adaptation performance over LTE for video streaming.

(a) Scenario without adaptation.

(b) Scenario with adaptation.

video streaming applications comes on the expense of the other running applications specially with the limited memory in the UE. The study states that given a network model characterized by the packet loss rate ($p$), RTT ($R$), and retransmission timeout ($T_0$), the receiver buffer size ($q_0$) that achieves desired buffer under-run probability ($P_u$) is given by

$$q_0 \geq \frac{0.16}{pP_u}[1 + \frac{9.4T_0^2}{bR^2}\min(1, 3\sqrt{\frac{3bp}{8}})p(1+32p^2)]. \quad (7)$$

where $b$ is the number of acknowledged packets. Another enhancement is suggested in [55] to enhance the TCP performance for video streaming over LTE by using the forward admission control to reduce the impact of handover between small cells on the TCP throughput. However, this scheme can be bandwidth consuming due to the extra signaling.

In general, TCP can be used for video streaming over LTE cellular network due to its highly preferable reliability. However, TCP lacks for satisfying the delay requirements due to the retransmission mechanism used in the TCP. Also, TCP can not guarantee the real time delivery for real time streaming services' packets such as video conferencing. Another disadvantage in TCP is that it stalls when packet losses happen.

### B. Real-time Streaming Protocol (RTSP)

RTSP is a protocol designed for entertainment and communication to control streaming media servers and facilitate media streaming over the network. RTSP provides several commands for the streaming such as pause, play, record and many more. RTSP can be used along with other protocols like TCP and UDP to add the urgency of time to the streaming data. RTSP can stream concurrent sessions and it keeps track of the state of each session. Furthermore, RTSP provides speed over reliability by using asynchronous QoS metrics such as packet-loss counts, jitter, and round-trip delay times [56]. RTSP can be used for real time traffic such as multi-player gaming and video calling such as Skype. RTSP is one of the main multimedia streaming protocols for the 3G mobile technology as well [57]. Hence, LTE can use similar standard as illustrated in [57] for live videos streaming. In [58], an analysis is conducted for the RTSP performance over LTE and WiMAX. The simulations in this study is done using OPNET [59] to measure various network statistics such as end-to-end delay, traffic throughput, jitter, and packet-loss. The RTSP ensures the packets to be delivered on time with a price of a higher packet losses or errors probability than TCP. The main disadvantage of the RTSP that it uses multi-cast that is not supported by many routers and occasionally is bing blocked by firewalls.

### C. Stream Control Transmission Protocol (SCTP)

This protocol was defined in 2000 by the Internet Engineering Task Force (IETF) [60]. SCTP is a message oriented transport protocol unlike TCP that transports a continuous data stream. One of the main advantages of SCTP is the multi-stream capabilities. This can help multi-interface devices (such as phones with WiFi and LTE) to receive video packets on different internet paths leading to a better video quality. This is beyond the scope of this survey, but more information for interested readers can be found in [61]. Another advantage to SCTP is the independent ordering for packets in each stream, which allows the application to choose processing the received messages by the order of receiving or the order they were sent in [62]. These two advantages makes SCTP a very desirable protocol for LTE traffic in general and multimedia streaming in particular. As we mentioned before, that most of the LTE packets loss results from the wireless environment degradation. SCTP overcomes the problem of traffic stopping and resource wasting due to the LTE channel variations because of the multi-stream capabilities. When an error happens in one stream, it does not affect any of the other streams, hence packet delivery is not suspended. SCTP has an advanced congestion control mechanism than the TCP, which consists of three phases: slow-start, congestion avoidance, and fast retransmition. As a result, many researchers directed their efforts to investigate the possibility and suitability of using the SCTP in the LTE including the 3GPP group themselves. A comparison between the TCP and SCTP in LTE has been introduced in [63]. The comparison recommends that SCTP is more suitable for LTE than the TCP due to the nature of multi-streaming.

### D. Summary and Discussion

In this section, we discussed different well known transport protocols. Based on our evaluations, we think that the SCTP is by far the most suitable protocol for stored video streaming over LTE among the discussed protocols and the RSTP is more suitable for the live multimedia streaming. TCP is very popular, widely deployed and well researched, hence it is still being used for some of the video traffic. However, there is plenty of room for research and enhancements to optimize the performance of these protocols or introduce new ones that can achieve a breakthrough for multimedia streaming over LTE. The new solutions must have the ability to reduce jitters in the high data rates and accurate data ordering and segmenting. Moreover, the future solutions should introduce new and fast ways to handle the congestions and packet losses taking into consideration the wireless channel variability. Our research group believes that the cross layers design approach can introduce a new dimension and provide new options to enhance the performance as it will be discussed in Section VII. Also, we think that using multiple layer protocols (such as one for delay and other for delivery) is not the optimal solution as it introduces more delay and complexity to the network. Another approach is using QoS aware or context aware protocols that can optimize the performance not only according to the traffic but also according to the user's experience and the wireless environment such as "the over the" top approaches followed in [64].

## VI. MAC LAYER MECHANISMS FOR VIDEO STREAMING

The MAC layer is responsible of addressing, channel accessing mechanisms, and organizing the medium sharing among different users. In the wireless environment, the MAC layer provides power control, bandwidth assignment, interference reduction, and collision avoidance [65]. In high speed cellular networks, different types of services with different priorities are being requested by the UEs such as web browsing, voice over IP, video calling, downloading files and many more. Hence, another mission for the MAC layer is to provide packet delay assurance and handle the different traffic priorities. For example, downloading a file have an overall rate and reliability requirements while video streaming or video conferencing have a more strict packet to packet delay requirements. That is to say, if a video frame is received late, it will be dropped as it is not needed anymore which will affect the video quality. The controls in the MAC layer is represented by the RLC loop and scheduler loop as shown in Figure 2. In the following subsections, we will discuss the different MAC layer aspects and its importance to the video delivery.

### A. Multiple Access Mechanisms

There are different common access protocols for packet wireless networks such as Carrier sense multiple access with collision avoidance (CSMA/CA) which is used in WiFi 802.11 [66]. Another common protocol is Code division multiple access (CDMA), where several UEs can simultaneously transmit over a single communication channel [67]. CDMA is commonly used in the UMTS. LTE uses OFDMA as the multiple access technique in the downlink and the Single-carrier FDMA (SC-FDMA) in the uplink. The SCFDMA has a lower Peak to Average Power Ratio (PAPR) than OFDMA which makes it favorable in the uplink to increase the UE power efficiency and battery life [68].

### B. Modulation and Coding Schemes

LTE uses the concept of adaptive modulation and coding (AMC) to change the packets coding and modulation based on the UE channel quality and the radio bearer requirements. These requirements depend mainly on the traffic type and is specified by the QoS Class Indicator (QCI) table given in the 3GPP standard [69] as shown in Figure 7. The QCI attribute determines the radio bearer traffic type, maximum allowed packet error and maximum packet delay. The AMC can be used to achieve these QCI and the rate requirements specially for video traffic. For example, a user with a strong channel can tolerate more errors, hence eNB can increase its overall rate by increasing the coding rate and increase the constellation order (bits/symbol) which will eventually lead to increasing the video quality by receiving more enhancement layers successfully. On the other hand, a user with a weak channel can not tolerate many errors and the eNB decides to use lower coding rates, more redundancy, which will eventually help decreasing the packet loss. This is essential in error sensitive applications such as interactive gaming or multi resolution broadcast as proposed in [70] and [71]. Hence, the eNB includes information in each packet that specifies the Modulation Coding Scheme (MCS) for the next packet.

### C. PRB Scheduling

The MAC layer in LTE is also responsible of managing the allocation functions, prioritizing the logical channel and its mapping to transport channels, scheduling information reporting, and managing HARQ, which is a transport-block level automatic retry. The MAC layer also selects transport format and provides measurements information about the network, while the radio link control (RLC) layer is responsible of packet segmentation and reassembly.

The LTE TTI scheduler is one of the vital MAC layer functions that has a great influence on the video quality over LTE. The time granularity of the TTI scheduler is the PRB unit i.e., 1 ms. The type of used scheduler by the eNB determine the resource distribution among the different users, hence the quality. The LTE scheduler has to ensure the time as well as the rate constraints for each user and radio bearer. One user can have multiple radio bearers, each carries different traffic type with different traffic constraints. Hence, researchers developed different types of LTE schedulers to cover many performance aspects. Some schedulers focus on enhancing the resource utilization and overall rate on the expense of fairness among users while others consider fairness as their first priority [72]. In the following, we will introduce some of the common TTI schedulers and discuss the trade-off between fairness and spectral efficiency. We also show how this trade-off affects the video traffic and quality. Scheduler algorithms in general can be categorized based on the channel knowledge into types; a channel non-aware and aware schedulers.

| QCI | Resource Type | Priority | Packet Delay Budget | Packet Error Loss Rate | Example Services |
|---|---|---|---|---|---|
| 1 | GBR | 2 | 100ms | 10-2 | Conversational voice |
| 2 | | 4 | 150ms | 10-3 | Conversational video (live streaming) |
| 3 | | 3 | 50ms | 10-3 | Real-time gaming |
| 4 | | 5 | 300ms | 10-5 | Non-conversation video (buffered streaming) |
| 5 | Non-GBR | 1 | 100ms | 10-3 | IMS signaling |
| 6 | | 6 | 300ms | 10-5 | Video (buffered streaming) TCP-based (e.g., www, email, chat, FTP P2P file sharing, progressive video, etc.) |
| 7 | | 7 | 100ms | 10-5 | Voice, video (live streaming), interactive gaming |
| 8 | | 8 | 300ms | 10-3 | Video (buffered streaming) TCP-based (e.g., www, email, chat, FTP P2P file sharing, progressive video, etc.) |
| 9 | | 9 | 300ms | 10-5 | |

Fig. 7: LTE standardized QCI values [69].

### D. Channel Non-Aware Schedulers

First in First out (FIFO) is considered the simplest channel unaware scheduler, however it is not efficient nor fair specially for the video traffic due to the frame delay limitations that varies between the different videos, services, and encoding. Another simple scheduler is round robin (RR) which guarantees the resource time occupation fairness but not the throughput fairness which is more important to the video traffic [73]. Weighted fair queuing (WFQ) identifies weights for each user/class of users based on their traffic type. This idea can help prioritizing traffic [74], however it needs to be used with another scheduler such as Round Robin to avoid starvation. Our research group thinks that the weights in the WFQ can be optimized as a function of the QCI values supported in the LTE MAC layer. A similar idea has been introduced using the packet delay instead of the Queue using the packet deadline metric and giving the highest priority to the packets with the closest deadline. This scheduler can be used in services such as video conferencing, where the quality is highly affected by the delay of the frames. A performance evaluation for the different channel non aware schedulers is conducted using OPNET in [75].

### E. Channel Aware Schedulers

The other type of LTE schedulers are the channel aware schedulers. There has been plenty of research to estimate the LTE channel such as [76]–[78]. Channel aware schedulers use the channel quality indicator (CQI) sent from the UEs to the eNB to estimate the channel quality between the eNB and the UE. The simplest channel aware scheduler is the maximum throughput which assigns the PRB to the UE with the maximum achievable throughput. Although this strategy can increase the UE's video quality, it does not take fairness into consideration [79]. Hence, starvation can happen to some users, moreover applications such as interactive gaming or video conferencing can be greatly degraded by the starvation. Proportional fair scheduler (PF) addresses the trade-off between the achievable throughput and the fairness. The main idea of the PF scheduler that it uses the average past received throughput as a metric which ensures that users with bad channel condition does not starve [80], [81]. Another channel aware scheduler is time to average throughput (TTA). The TTA can be considered as an intermediate scheduler between the maximum throughput and the PF. The metric for this scheduler is the maximum throughput for a certain PRB normalized to the overall average throughput. The more average throughput a user can obtain, the lower the metric is. Hence, it gives an opportunity to other users to utilize the resources and to decrease the probability of starvation. However, this metric does not take the time constraints into consideration which makes it not suitable for delay sensitive videos. A performance evaluation has been carried out via NS3 [82] in [83] to compare between the different schedulers implementation in the LTE environment.

### F. Summary and Discussion

LTE MAC layer has many enhancements and advantages that can accommodate the streaming of high quality and high definition videos over the cellular networks and adapt the effect of the channel change in the channel environment. However, we believe that there is still a lot of space to add more enhancements and introduce new algorithms that is specifically designed for video streaming capabilities. These new algorithms shall integrate the advantages of the MAC layer such as AMC and the new proposed LTE scheduler with the video quality of service metrics and encoding information used by the service provider to create a hybrid/cross layer design schemes for multimedia applications as it will be discussed in Section VII. Some open research questions in this area is which metrics should be used and what are the effects for partial and/or full use of all the available metrics and feedback information. Furthermore, the complexity and the processing overhead in the MAC layer for these hybrid algorithms needs to be studied to evaluate the quality gain versus the complexity.

## VII. CROSS LAYER TECHNIQUES FOR MULTIMEDIA STREAMING APPLICATIONS

Based on the earlier discussion regarding the LTE cellular network and the work done to accommodate multimedia streaming applications, it clearly appears that the performance of LTE cellular networks is not yet optimized for end to end multimedia streaming delivery. This is normal as LTE is not designed to transmit video traffic alone but also for other types of services such as web surfing and file transfer in addition to voice calls using the Vo-LTE. The previously discussed delivery algorithms and mechanisms are limited with the LTE standard and optimizes only in one of the seven layer of the Open Systems Interconnection OSI model introduced in [84].

The OSI layered model forces the hierarchy among the layers and does not allow communication between them, however this defies the dynamic nature of the new cellular networks including LTE. In other words, the new cellular environments with all their different capabilities, multiple equipments capabilities and dynamic traffic need to be self adapting based on the dynamic factors in the network. Such concept of self configurable heterogeneous networks is hard to achieve without exchanging information between the network layers, i.e. cross layer design [85]. Moreover, to optimize the end to end delivery performance for any application in general and video streaming over the LTE network in particular, application and network information need to be exchanged between the different layers. Hence, in cross layer design we can exploit the layers dependency instead of treating each layer as an independent entity [5]. That is to say, the control loops illustrated in Figure 2 are either exchanging information or merging together. This helps optimizing the entire network elements performance at once, while exploiting the ability of the LTE radio network to early detect variations. It is important to note that the LTE network time granularity is much faster (order of milliseconds) than the end to end time frame. For the rest of the section, we will discuss some of the recently proposed solutions that involves cross layer design to optimize multimedia applications delivery over wireless cellular networks in general and over LTE in particular.

Different ideas have been proposed to optimize the video transmission over LTE. Most of these ideas depend mainly on changing the video and network parameters based on the environment, channel quality and video packets information and priority. For example, a game theoretic spectrum agility approach is used in [86] to ensure the delivery of sensitive delay applications over wireless networks. The goal of this work is to maximize the number of satisfied users while ensuring a fast reaction for secondary users in a cognitive radio network which is also known as Opportunistic Spectrum Agile Radios (OSAR). This is achieved by sharing the desired video QoS information among the different users to be considered in the scheduling phase. A multi description coding is used in [87] to change the 802.11 MAC layer to be adaptive to the wireless channel so that the receiver can change its received quality by receiving more descriptions of the video frames when it has a good link quality. Similar ideas have been proposed in general wireless networks and based on the cross layer over wireless network framework described in [88] have inspired the cross layer design for video delivery over cellular networks in general and LTE in particular.

Our research group have some work in cross layer design over LTE. In this work, the mutual information between the video layers is the context. In [23], we propose a content aware scheduler for the MPEG4 video frames over WCDMA cellular networks. In this work, we mutually design the application layer and the MAC layer, where the transmitter decides which enhancements frames to add in a group of pictures structure based on the channel quality between the eNB and the UE. In addition to working with WCDMA, our group proposed a novel LTE scheduler in [89] to optimize H.264 videos delivery over LTE. In this work, we also mutually design the MAC and the application layers to simultaneously choose the number of enhancement levels of an H.264 encoded video that should be transmitted by the content provider for each user to maximize the quality over the network. The eNB is also considered content aware in this scenario, where it assigns more PRBs to the users who have higher priorities videos, or videos need high quality, such as action scenes, or with a low channel quality to ensure no starvation while maximizing the QoS across all users. A similar work is done by our collaborators in [90] to optimize the delivery of DASH videos over LTE cellular network. We extended our previous work to include a cross layer design between the MAC, TCP and the application layer in [91] by making the LTE scheduler aware to the UE and eNB buffers status in addition to being originally content aware. Hence, a scheduler can effectively assign resources to the users in need without video freeze or buffer overload.

A merge between the MAC and RTP is proposed in [92] where the eNB MAC layer sends the average CQI information per user to the video RTP server. The video RTP server decides which temporal enhancement layers to drop based on a predefined look-up table. This scheme shows 30% enhancement in the quality of low CQI users as shown in Figure 8. A similar work is also proposed in [93]. In [94], a MOS-based QoE predictionfunction is derived to maximize the users quality while guaranteeing fairness among them. This research also establishes a mapping between PSNR and the user's opinion based on a tangent function curve to outline analytically the relation between the PSNR and a human visual perceiving model. Finally, it adopts the Particle Swarm Optimization (PSO) [95] to find the optimal resource allocation based on the quality of experience of each user.

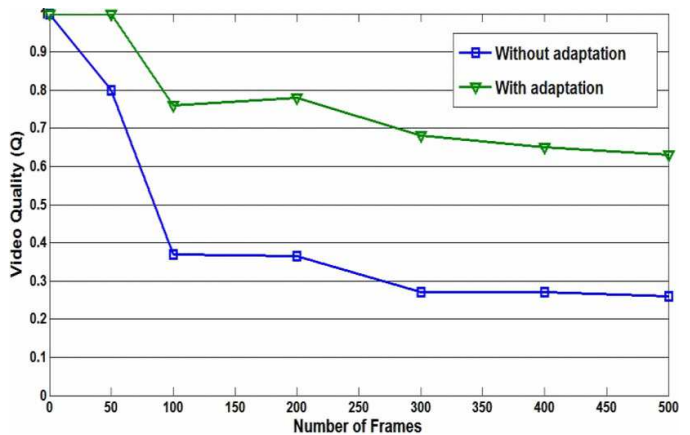

Fig. 8: Video quality of low CQI user (CQI 3 and 4) with and without adaptation [92].

## A. Summary and Discussion

As we have seen in the previously discussed schemes, cross layer designing can enhance the performance because it dynamically adapts the parameters of different network layers simultaneously based on the video information and the quality of the link between the UE and the eNB. However, this comes with some trade-off and limitations. In this subsection, we will discuss some of the main limitations in the cross layer schemes and propose some of the open research problems in this area. The first limitation is that most of the proposed algorithms in this area are centralized. Centralized algorithms usually come with a high computational complexity, intensive signaling and violation for the standard rules in expense of the performance. It is rare to find a centralized approach, such as our work in [89], [91], that can be a plug and play without much modifications in the network or signaling overhead on the radio core. A complex centralized scheme is not likely to be commercially implemented rather than just clarifying that cross layer design can significantly enhance the video quality. Hence, we think that developing distributed cross layer schemes is an open research problem and also critical to be able to implement in reality. The work in [96], [97] can be considered a paradigm for distributed cross layer optimization schemes. The proposed scheme in [96] suggests that each user will run an optimization problem to determine the number of resources needed to satisfy certain video delay requirements given the number of users in each network in a multiple heterogeneous network systems. Similar schemes need to be studied over LTE networks in addition to complying with the standard. Furthermore, there is an extra overhead, whether the scheme is centralized or distributed. This overhead is introduced by the signaling between the layers and the different users to collect their channel and video information. Hence, studying the signaling and performance overhead is another open problem. Signaling can also be a serious problem specially during the UE handoff. Handover in LTE happens frequently due to supporting high speed mobility. When a handover happens, the UE is added to a new set of users and has to exchange information again as the previous information is useless. Hence, this extra signaling is considered a waste during the handoff time as well as the old information.

Another open research question is determining the helpful attributes of the video application to include in the optimization problem along with the network information. These attributes can range from high level such as genre (For example, action movies in general requires higher bit rate than musical video clips to achieve the same quality), or the application type. It can also be in a microscopic level such as frame delay, and coding settings. Hence, it is vital to analytically quantify and experimentally test the contribution of different subsets of these attributes to the user's experience while simultaneously reducing the signaling and complexity overhead. Moreover, the lack of unified simulation model as we mentioned earlier makes the comparison between schemes very hard as some of the cross layer schemes sacrifices the reality of the model to show significance improve in the quality. However, It becomes hard to check this and compare between the multiple existing schemes with the lack of incorporated testing model or a quantitative analysis for the proposed schemes.

There exist some scenarios where information exchange and signaling for the cross layer optimization is desirable and helpful rather than being overhead. According to [98], the resources are shared in the eMBMS mode and the MBMS bearer service uses IP multi-casting to deliver its traffic. The core network can decide to assign some users more uni-cast resources to individually enhance their quality if possible. To the latest information of the authors, there has not been a solid work to exploit the advantages of the cross layer optimization in the eMBMS over the LTE networks.

Finally, we think that finding implementable cross layer techniques that takes into consideration other traffic types beside video application traffic is an important research topic. As we explained before, LTE can carry different types of traffic and support a lot of different applications due to it's high data rate and mobility support. Hence, it is important to ensure that the video traffic does not compromise the performance of other applications. Hence, an implementable proper cross layer design must consider other traffic types and their priorities.

## VIII. CONCLUSION

In this survey, we discussed different optimization aspects over cellular networks, in particular LTE, in order to enhance the delivery of the video streaming services to the end user. We discussed the different metrics that can be used to characterize the performance of proposed solutions. In addition, we highlighted the limitations of a per layer basis solutions and pointed out to the environment and assumptions accompanied to these solutions in order to perform well. In our opinion, we think that cross layer techniques lead to better end to end optimization and takes into consideration a lot more optimization parameters compared to the per layer optimization techniques.